\documentclass[aps,prf,reprint,groupedaddress,longbibliography]{revtex4-1}
\usepackage{graphicx}
\usepackage{natbib}
\usepackage{amsmath,amssymb}
\usepackage{subfig}
\usepackage{setspace}
\usepackage{float}

\usepackage[usenames, dvipsnames]{color}
\usepackage{hyperref}
\hypersetup{
	hyperindex,
	breaklinks,
	colorlinks=true,
	linkcolor=blue,
	citecolor=magenta,
	bookmarks=true,
	bookmarksopen=true,
	bookmarksopenlevel=2,
	pdfstartpage={1},
	pdfstartview={FitH},
	pdfview={FitH 0},
	pdfauthor={B. F. Farrell and P. J. Ioannou},
	pdftitle={}}

\usepackage{ifthen}

\def\U{\bm{\mathsf{U}}}
\def\Uv{\boldsymbol{U}}
\def\uv{\boldsymbol{u}}

\def\U{\bm{\mathsf{U}}}

\def\Uv{\boldsymbol{U}}

\newcommand{\be}{\begin{equation}}
\newcommand{\ee}{\end{equation}}
\newcommand{\bdm}{\begin{equation*}}
\newcommand{\edm}{\end{equation*}}
\newcommand{\bea}{\begin{eqnarray}}
\newcommand{\eea}{\end{eqnarray}}

\newcommand{\partialf}[2]
{
 \ifthenelse{\equal{#1}{}}{\frac{\partial}{\partial #2}}{\frac{\partial #1}{\partial #2}}
}

\newcommand{\real}{\mathop{\mathrm{Re}}}

\renewcommand{\(}{\left(}
\renewcommand{\)}{\right)}
\renewcommand{\[}{\left[}
\renewcommand{\]}{\right]}

\newcommand{\upi}{\pi}

\newcommand{\df}{\textrm{d}}

\renewcommand{\i}{i}

\newcounter{saveeqn}%

\def\bit{\vphantom{\dot{W}}}

\def\Uv{\boldsymbol{U}}
\def\uv{\boldsymbol{u}}

\newcommand{\defn}{\ensuremath{\stackrel{\mathrm{def}}{=}}}
\renewcommand{\equiv}{\defn}

\def\bit{\vphantom{\dot{W}}}

\renewcommand{\U}{\boldsymbol{U}}
\renewcommand{\u}{\boldsymbol{u}}

\begin{document}

\title{Instability of the roll/streak structure induced by free-stream turbulence in pre-transitional Couette flow}

\author{Brian F. Farrell}
\affiliation{School of Engineering and Applied Science, Harvard University}
\author{Petros J. Ioannou}
\email{pjioannou@phys.uoa.gr}
\affiliation{Department of Physics, National and Kapodistrian University of Athens}
\author{Marios-Andreas~Nikolaidis}
\affiliation{Department of Physics, National and Kapodistrian University of Athens}
%

\date{\today}

\begin{abstract}
Although the roll/streak structure is ubiquitous in both observations and simulations of pre-transitional
wall-bounded shear flow, this structure is linearly stable if the idealization of laminar flow is made.  Lacking an
instability, the large transient growth of the roll/streak structure has been invoked to explain its appearance as  resulting from
chance occurrence in the background turbulence of perturbations configured to optimally excite it.
However, there is an alternative interpretation for the role of free-stream turbulence in the genesis of the roll/streak structure
which is that the  background turbulence interacts with the roll/streak structure to destabilize it.
Statistical state dynamics (SSD)
provides analysis methods  for studying instabilities of this type which arise from interaction between the coherent and incoherent components of turbulence.
Stochastic structural stability theory (S3T), which implements SSD in the form of a closure at second order,
is used in this work to analyze the SSD modes arising from interaction between  the coherent streamwise invariant  component   and the incoherent  background  component of turbulence.
In pre-transitional Couette flow a manifold of stable modes with roll/streak form is found to exist in the presence of  low intensity background turbulence.  The least stable mode of this manifold is destabilized at a critical value of a parameter controlling the background turbulence intensity and a finite amplitude roll/streak structure arises from
this instability through  a  bifurcation in this parameter.  Although this bifurcation has analytical expression only in SSD,  it
is closely reflected in both the dynamically similar quasi-linear system, referred to as the restricted non-linear (RNL) system, and in
DNS.  This correspondence is verified using ensemble implementations of the RNL and DNS systems.    S3T also predicts
a second bifurcation at a higher value of the turbulent excitation parameter that results in destabilization of the finite amplitude
roll/streak equilibria.  This second bifurcation is shown to lead first to time dependence of the roll/streak in the S3T system
and then to chaotic fluctuation corresponding  to minimal channel turbulence.  This
transition scenario  is also verified  in simulations 
of the RNL and DNS systems.   Bifurcation from
a finite amplitude roll/streak equilibrium provides a direct route  to the turbulent state through the S3T roll/streak instability.
 \end{abstract}


\maketitle

\section{Introduction}

Streamwise roll vortices and associated streamwise streaks were identified in experiments on transition   in boundary layers ~\citep{Klebanoff-etal-1962} and observed in the near wall region of turbulent flows  ~\citep{Kline-etal-1967, Blakewell-Lumley-1967, Kim-Kline-1971}.  These observations were subsequently corroborated by direct numerical simulations (DNS)  (cf.~\citet{Kim-etal-1987}) and the roll/streak structure  is now understood to be central to the dynamics of turbulence in wall-bounded shear flows.

There are two distinct dynamical problems central to understanding wall-turbulence: transition from the laminar to the turbulent state and maintenance of the turbulent state.  The roll/streak structure, despite being hydrodynamically stable,  is commonly agreed to be involved in instigating  transition from the laminar to the turbulent state in these flows.  After transition this structure persists  but becomes  highly variable in both space and time.  This time-dependent
streamwise roll and streak structure is believed to be  involved in the process maintaining  turbulence in shear flow
that is referred to as the self-sustaining process \citep{Jimenez-Moin-1991, Hamilton-etal-1995, Schoppa-Hussain-2002, Jimenez-2013}.    Moreover, this self-sustaining mechanism appears to be quite general in that it operates not only in the near-wall region but also, and independently, in the logarithmic layer \citep{Hwang-Cossu-2011, Farrell-etal-2016-VLSM}.

Our primary interest in this work is in the robust observation of  the roll/streak  structure in wall-bounded shear flow prior  to transition
and in understanding the role of this structure in the transition process.  The prominence
of the roll/streak in these flows presents a problem because this structure is not an unstable eigenmode of the shear flow
existing prior to transition. The robust
observation of the roll/streak structure was first rationalized by appeal to the lift-up mechanism which describes the kinematic conversion of wall normal velocity into streamwise streak velocity in sheared flows \citep{Ellingsen-Palm-1975, Landahl-1980}.
This insight was later advanced by recognition that the lift-up mechanism could be subsumed into the analytical structure of generalized stability theory (GST) by which modal stability theory and non-normal transient growth analysis are united  \citep{Farrell-Ioannou-1996a,
Farrell-Ioannou-1996b,Schmid-Henningson-2001}.  While modal stability analysis provides no reason to expect appearance of roll/streak structures,  GST analysis predicts optimally growing perturbations with the observed form \citep{Butler-Farrell-1992, Reddy-Henningson-1993}.

The success of optimal growth theory in predicting the  roll/streak structure observed in
perturbed  wall-bounded shear flow prior to transition appeared at first to be persuasive that the explanation for
observations of this structure in pre-transitional flow was secure.
Nevertheless, there remained a lingering doubt.  For one thing, there is the regularity of the spacing and amplitude of the roll/streak in experiments \citep{Head-Rechenberg-1962, Bradshaw-1965},
which, as remarked by Townsend \citep{Townsend}, is characteristic of modal growth.  And then there is the  observation that streamwise rolls decay in amplitude if background turbulence levels are sufficiently low, consistent with predictions based on transiently growing optimals \citep{Bakchinov-etal-1997, Alfredsson-Matsubara-1996, Westin-etal-1998}, while rolls grow downstream in the presence of moderate levels of background turbulence  intensity \citep{Westin-etal-1994},  which is incompatible with transient growth
and suggestive of  an underlying unstable mode.

While the absence of roll/streak  instability in an unperturbed wall-bounded shear
flow is established,  pseudospectral theory \citep{Trefethen-etal-1993,Trefethen-2005} reveals that a highly non-normal operator, such as that of Navier-Stokes  (NS)
dynamics linearized about a strongly sheared flow, can be destabilized by small perturbations to the dynamical operator itself.   Consistently, it was recently shown that an emergent instability with roll/streak structure  arises from interaction between the roll/streak structure and
a field of background turbulence with sufficient amplitude \citep{Farrell-Ioannou-2012}.
This  instability does not have
analytical expression in the linearized NS dynamics of the laminar flow because it is not a linear  instability of the laminar
shear flow  but instead arises from systematic organization
by the roll/streak structure of the Reynolds stress associated with the incoherent background turbulence.
The analytical expression for this instability therefore exists only in the equations for the associated statistical state  dynamics (SSD).
The formulation of SSD used in this work  to study this instability, referred to as S3T,  is a second order closure of the Navier-Stokes
dynamics (NS)
in which full nonlinearity is retained in the streamwise mean equation (first cumulant)  while the  dynamics of the perturbation covariance  (second cumulant)
is linearized about the instantaneous streamwise mean flow.
Nonlinear interaction occurs between the mean flow dynamics (defined as flow components with streamwise
wavenumber $k_x=0$) and the perturbation covariance obtained from flow components with streamwise wavenumber
$k_x\ne 0$, while nonlinearity is parameterized by a stochastic excitation in the perturbation dynamics rather than being explicitly calculated.
This quasi-linear formulation in which nonlinearity is parameterized in the perturbation dynamics is referred to as the restricted
nonlinear (RNL) approximation to the full nonlinear Navier--Stokes  dynamics (NL).
In this work we use RNL to construct finite ensemble approximations to the equivalently infinite ensemble of
the S3T version of SSD.  Consistent with this usage, the perturbation equations making up the ensemble in an RNL-based
approximation to S3T are used only to calculate an approximate covariance.
As a consequence  phase information is not retained for the perturbation fields, only
their second order correlations being relevant to a second order SSD.

As alluded to above, the approximation to the perturbation covariance
obtained using RNL dynamics can be systematically improved by forming a mean covariance
from an ensemble of RNL perturbation equations sharing a single mean flow.   In the case that an $N$-member ensemble is used to approximate the covariance the SSD approximation is referred to as
RNL$_N$  \citep{Constantinou-etal-2016}.  In the limit $N \rightarrow \infty$ S3T dynamics is recovered.
RNL has the advantage that it can be easily implemented at high resolution while retaining the analytical restrictions of S3T.  Moreover,  simulations made using RNL can be compared to the same DNS implementation that was restricted to obtain the RNL system~\citep{Thomas-etal-2014,Farrell-etal-2016-VLSM}.

Further insight can be obtained by proceeding similarly with the NS equations by formally writing the full
dynamics in mean/perturbation form and then calculating an ensemble average second order
closure using an $N$-member ensemble of perturbation equations sharing a single mean flow in a manner
parallel to the method used in constructing RNL$_N$ but retaining full nonlinearity in the individual perturbation equations of the ensemble.  This closure will be referred to as NL$_N$.  When it converges NL$_N$ corresponds to a complete cumulant expansion of the SSD solved up to second order.
We find that in our example problem satisfactory convergence of RNL$_N$ and NL$_N$ is obtained for $N$ as small as $10$.


Consider a Couette flow  subjected to a  random excitation that is  statistically streamwise and spanwise homogeneous and has zero mean  with respect to time  and  space averaging.   S3T predicts a bifurcation occurring at a critical amplitude of the excitation in
which an unstable  roll/streak structure emerges as an instability of the S3T dynamics.  {It is important at this point to be clear about  what entity is being referred to as unstable.  The unstable mode we are studying
arises as an eigenmode with roll/streak structure at infinitesimal amplitude  that eventually grows sufficiently to
become a nonlinearly equilibrated finite amplitude equilibrium that retains roll/streak structure. The existence of
coherent roll/streak structures in the flow is therefore explained by the growth and equilibration
of this unstable mode. It is perhaps more correct to say that the flow is unstable “to” this roll/streak
structure than to say that this roll/streak structure is unstable, which would admit
the alternative  interpretation that the finite amplitude roll/streak structure is itself unstable.  At sufficiently high
background turbulence levels the finite amplitude roll/streak structure proceeding from the S3T unstable mode does become
itself subject to secondary instability leading to transition to a self-sustaining turbulent state as we will show.
The perturbative  S3T instability
connects directly to the finite amplitude roll/streak structure which becomes secondarily unstable,
but these secondary instabilities are not of roll/streak form.
There is an analogy between equilibrated finite amplitude roll/streak structures in S3T and  exact coherent structures  in laminar flow
\citep{Nagata-1990,Hall-Smith-1991,Waleffe-2003,Wang-etal-2007,Deguchi-etal-2013, Deguchi-Hall-2014}, although exact coherent
structures  are finite amplitude isolated
equilibria that do not connect to infinitesimal instabilities of the  spanwise independent laminar flow  as the S3T roll/streak structures do.   While S3T finite amplitude roll/streak structures become secondarily unstable only when these roll/streak reaches high amplitude under excitation by strong background turbulence, the isolated exact coherent structures generally  support secondary instabilities, for example
those discussed  in Deguchi \& Hall
\cite{Deguchi-Hall-2014,Deguchi-Hall-2016} in their investigation of the stability of the finite amplitude states  in
vortex-wave interaction theory (VWI).
 We remark that once its secondary instability becomes supported the coherent equilibrium S3T roll/streak structure is rapidly  destroyed.  This observation suggests that physically realistic levels of background turbulence should excite the parasitic modes of exact coherent structures  as well.  In order to maintain such unstable structures it is necessary to eliminate
naturally occurring sources of perturbations that would necessarily excite the parasitic modes to
which these structures are vulnerable. In contrast, the S3T instability results from organization of
the background disturbances which constitutes its energy source so rather than being detrimental
to it, the S3T mode growth rate increases with increasing background disturbance amplitude.}

 Returning now to the S3T instabilities with roll/streak form; as the background turbulence  excitation is increased, at
 first  the streamwise and spanwise averaged mean flow
differs little from the laminar Couette profile while superimposed on this profile is a fixed point finite amplitude roll/streak structure.
With further increase in the excitation amplitude  another critical  value is exceeded at which the flow transitions to turbulence. The existence of
these  three statistical regimes under increasing levels of background turbulence:
the near laminar state, the near laminar with superimposed finite amplitude equilibrated roll/streak structure,  and the turbulent regime characterized by  chaotic fluctuation of the roll/streak structure in Couette flow was predicted  using S3T  \citep{Farrell-Ioannou-2012}.
%
The purpose of this paper is to determine whether these predictions made using S3T are  reflected in ensemble  RNL and NL SSD approximations
and to analyze the convergence to the  S3T predictions obtained using the RNL$_N$ and NL$_N$ approximations as $N \rightarrow \infty$.

\section{Formulation of S3T }

\label{sec:framework}
\begin{figure*}
	\centering
       \includegraphics[width = .8\textwidth]{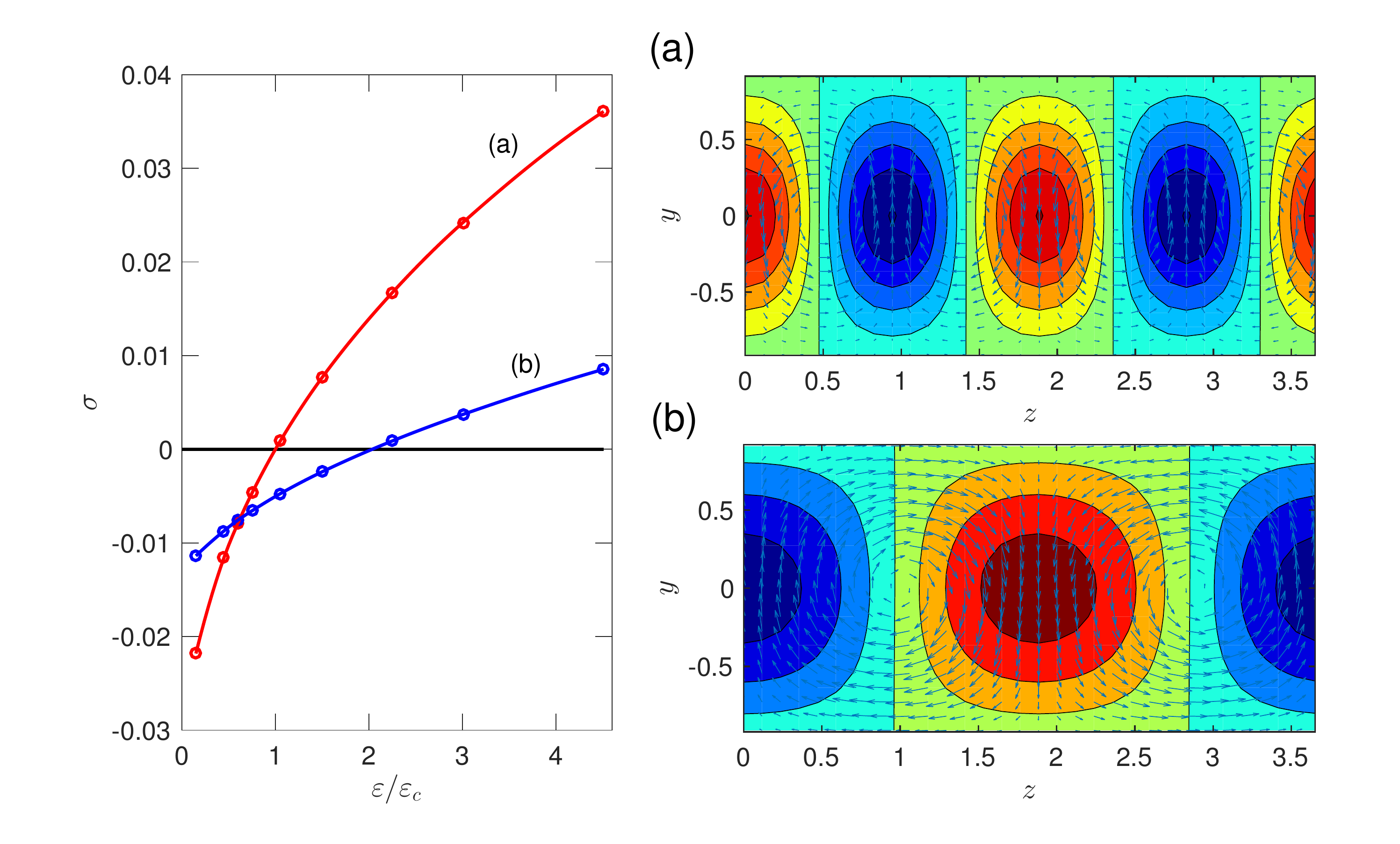}
        \caption{ Left: Growth rate of the two most unstable  S3T eigenfunctions about the spanwise homogeneous S3T equilibrium as a function of the excitation amplitude  of the background turbulence,
$\varepsilon$.   Right: The structure of the corresponding eigenfunctions with growth rate (a) and (b) for excitation amplitude  $\varepsilon / \varepsilon_c=2$.
Shown are contours of the streak velocity, $U_s$, and velocity vectors of the components $(V,W)$  plotted on a $(y,z)$ plane cross-section.
The structure of these  eigenfunctions  does  not change appreciably for $\varepsilon/\varepsilon_c  < 6$.
At  $\varepsilon=\varepsilon_c$  the S3T spanwise uniform equilibrium bifurcates to a finite amplitude equilibrium
with perturbation structure close to that of the most unstable eigenfunction  shown in (a).
The channel is minimal with $L_x=1.75 \upi$ and $L_z=1.2 \upi$,
the Reynolds number is $R=400$, and the stochastic forcing excites only  Fourier components with  streamwise wavenumber $k_x= 2 \upi/L_z= 1.143$.
The critical $\varepsilon_c$  sustains a background turbulent  field with mean energy $0.14 \% $ of the Couette flow energy.}
\label{fig:grS3T}
 \end{figure*}


Consider a plane Couette flow with streamwise direction $x$, wall-normal direction $y$ and spanwise direction $z$  in which
background turbulence is  maintained by stochastic excitation applied throughout the flow.   The  lengths of the
channel in the streamwise, wall-normal and spanwise  direction are respectively $L_x$, $2h$ and $L_z$. The channel walls are at $y/h=-1$~and~$1$.  Spatial and temporal averages are denoted by  square brackets with a subscript denoting the independent variable over which the average is taken, i.e.~spanwise averages by $[\,\boldsymbol{\cdot}\,]_z=L_z^{-1} \int_0^{L_z} \boldsymbol{\cdot}\ \df z$, time averages by~$[\,\boldsymbol{\cdot}\,]_t=T^{-1} \int_0^{T} \boldsymbol{\cdot}\ \df t$, with  $T$ sufficiently long.
Multiple subscripts denote an average over the subscripted variables  in  the order they appear, i.e. $[\,\boldsymbol{\cdot}\,]_{x,y} \equiv \[\,[\,\boldsymbol{\cdot}\,]_{x}\,\]_{y}$.  The vector velocity $\uv$ is decomposed into its streamwise mean, denoted by $\Uv(y,z,t)\equiv\[\u(x,y,z,t)\]_x$, and the deviation from this mean (the perturbation) denoted $\u'(x,y,z,t)$ so that  $\uv = \Uv + \uv'$. The pressure gradient is similarly decomposed as $\nabla p= \nabla\( P(y,z,t)+ p'(x,y,z,t) \)$.  Velocity is  non-dimensionalized
by the velocity at the wall, $U_w$, at $y/h=1$, lengths by $h$, and time by $h/U_w$.   The non-dimensional NS equations decomposed into an equation for the  mean and an equation for the perturbation are:
\begin{subequations}
\label{eq:NS}
\begin{gather}
\partial_t\boldsymbol{U}+ \boldsymbol{U} \cdot \nabla \boldsymbol{U}   + \nabla P -  \Delta \boldsymbol{U}/R =- \[\boldsymbol{u}' \cdot \nabla \boldsymbol{u}'\]_x\ ,
\label{eq:NSm}\\
 \partial_t\boldsymbol{u}'+   \boldsymbol{U} \cdot \nabla \boldsymbol{u}' +
\boldsymbol{u}' \cdot \nabla \boldsymbol{U}  + \nabla p' -  \Delta  \boldsymbol{u}'/R
= \nonumber\\\qquad= - \(  \boldsymbol{u}' \cdot \nabla \boldsymbol{u}' - \[\boldsymbol{u}' \cdot \nabla \boldsymbol{u}'\]_x \,\)+ \sqrt{\varepsilon} ~{\boldsymbol{f}'}\ ,
 \label{eq:NSp}\\
 \nabla \cdot \boldsymbol{U} = 0\ ,\ \ \ \nabla \cdot \boldsymbol{u}' = 0\ ,\ \ \ \nabla \cdot \boldsymbol{f} = 0\, \label{eq:NSdiv0}
\end{gather}\label{eq:NSE0}\end{subequations}
where $R= U_w h/ \nu$ is the Reynolds number.
The velocities and the  stochastic excitation $\boldsymbol{f}'(x,y,z,t)$  satisfy periodic boundary conditions  in the $z$ and $x$ directions
and no-slip boundary conditions in the cross-stream direction: $\boldsymbol{U}(x,\pm 1,z,t)= (\pm 1,0,0)$, $\boldsymbol{u}'(x,\pm1,z,t)=\boldsymbol{f}'(x,\pm 1,z,t)=0$.
The stochastic excitation is applied  only to the streamwise varying Fourier components
of the flow.  It  is nondivergent, has zero ensemble mean,  $\left < \boldsymbol{f}' \right > =0 $ (the ensemble mean over excitation realizations being denoted $\left < \cdot \right >$) , and
is delta correlated in time and  statistically homogeneous  in the  $x$ and $z$ directions.
Delta correlation in time of the excitations implies that the energy input by the
stochastic excitation is independent of the flow state and can be  parameterized by $\varepsilon$ in (\ref{eq:NSp}).  
%
The $x,y,z$ components of $\U$ are $(U,V,W)$ and the corresponding components of $\u'$ are $(u',v',w')$.
The streak component of the
streamwise  mean flow is denoted by $U_s$ and defined as
\begin{equation}
U_s\equiv U-[U]_z ~.
\end{equation}
The streamwise mean cross-stream and spanwise velocities, $V$ and $W$,  are found  to primarily constitute the roll vortices.
We also define the streak  energy density, $E_s=\[U_s^{2}/2\]_{y,z}$,  the roll energy density,
$E_r= \[(V^2+W^2)/2\]_{y,z}$, and the perturbation energy density $E_p= \[ |\u'|^2/2 \]_{x,y,z}$.
Energy  is injected from the moving walls at  rate   $I= (2R)^{-1} \[ \left . \partial_y U \right|_{y=1}+\left .  \partial_y U \right |_{y=-1} \]_z$
and at  rate $\varepsilon$ from the  appropriately normalized stochastic forcing.
Energy  is dissipated at  rate $D=R^{-1} \[ |\nabla \times { \u }|^2 \]_{x,y,z}$. With $I_c$ and $D_c$ we denote the energy injection and
dissipation rates of the Couette flow.   

The S3T dynamics is a SSD  governing the evolution  of the first two cumulants consisting of the
streamwise mean flow,
$\U=(U,V,W)$  or $\U  \equiv (U_x,U_y,U_z)$,
and the second cumulants that are the same time covariances of the Fourier components of the  velocity fluctuations, $\hat{u}_{\alpha, k_x}'$,
where the index $\alpha=x,y,z$ indicates the velocity  component in the Fourier expansion
of the  perturbation velocity  $\uv'$:
\begin{equation}
\uv' (x,y,z,t)= \sum_{k_x>0} \real \left (  \hat{\uv}'_{k_x} (y,z,t)\,e^{\i k_x x} \right )\ ,
\end{equation}
with $k_x$ the streamwise wavenumbers that are excited by the stochastic excitation.  We similarly expand the excitation in its Fourier components
$\hat{\mathbf f}'_{k_x}$. In this study we will limit the
stochastic excitation to  only the streamwise fundamental wavenumber $k_x=2 \upi/L_x$ and as a result the subscript
$k_x$ in the velocity and excitation components can be dropped without ambiguity. Because  in the S3T equations the perturbation-perturbation interactions are not included,
this choice of excitation implies that the  S3T flow field  perturbations have power  only at the streamwise component that is forced.
 The covariance variables of S3T are the covariances of the velocity components
of  Fourier component $k_x$  between point
$1\equiv(y_1,z_2)$ and point $2\equiv (y_2,z_2)$  evaluated at the same time:
\begin{equation}
C_{\alpha \beta}(1,2)= \left < \hat{u}_{\alpha}'(1) \hat{u}_{\beta}'^*(2)  \right >~,
\end{equation}
which is a function of  the coordinates of the two points $(1)$ and $(2)$ on the $(y,z)$ plane and of time ($*$ denotes complex conjugation). The S3T equations are:
\begin{subequations}
\label{eq:S3T}
\begin{gather}
\partial_t U_\alpha+ U_\beta \partial_\beta U_\alpha + \partial_\alpha P - \Delta U_\alpha/R =\nonumber\\\hspace{4em}= -
\frac{1}{2}  \real \left (  \bit  \partial_y  C_{y \alpha} (1,1) + \partial_z  C_{z \alpha} (1,1)\right ) \ ,
\label{eq:S3Tm}\\
\partial_t C_{\alpha \beta}(1,2)= A_{\alpha \gamma}(1) C_{\gamma  \beta}(1,2) \nonumber\\\hspace{7em} + A_{ \beta \gamma}^*(2) C_{\alpha \gamma}(1,2) + Q_{\alpha \beta} (1,2) \ , \label{eq:S3Tp}\\
\partial_a U_a =0 ~~,~~\hat{\partial}_\alpha (1) C_{\alpha \beta} (1,2)= \hat{\partial}_\beta^*(2) C_{\alpha \beta} (1,2) = 0\ , \label{eq:S3Tdiv}
\end{gather}\label{eq:QLE0f}\end{subequations}
with summation convention on repeated indices and the operator $\hat{\partial} \equiv ( i k_x , \partial_y, \partial_z )$ (for a derivation cf. \cite{Farrell-Ioannou-2012}). The operator
$A_{\alpha \beta} (1)$ (or $A_{\alpha \beta} (2)$)  is the operator governing the quasi-linear  evolution of streamwise  varying perturbations in \eqref{eq:NSp} with streamwise wavenumber $k_x=2 \upi/L_x$
linearized about the instantaneous streamwise mean flow $\U(1)$ (or $\U(2)$) and $1$ (or $2$) indicates that the operator acts on the $1$ (or the $2$)
variable of $C(1,2)$.  $Q_{\alpha \beta}(1,2)$ are the spatial covariances of the $k_x$ Fourier components of the forcing components,  $\hat{f}_i$, and are defined as
\begin{equation}
\left < \hat{f}_\alpha(1,t_1) \hat{f}_\beta^*(2,t_2 )\right > = \delta(t_1-t_2) Q_{\alpha \beta}(1,2)~.
\end{equation}

\begin{figure*}
	\centering
       \includegraphics[width = .6\textwidth]{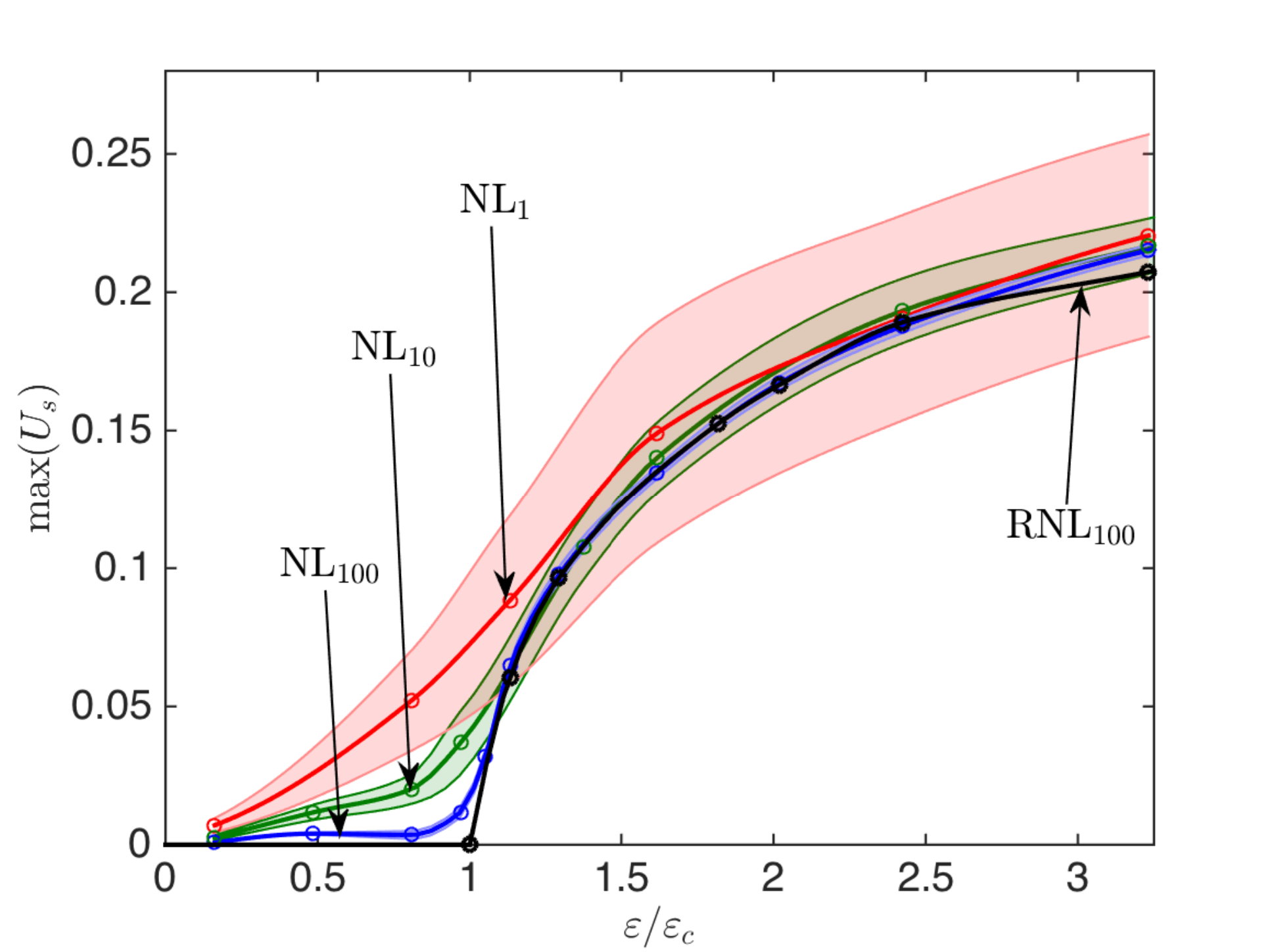}
        \caption{Bifurcation diagram for the Couette problem. Shown is
the time mean of the  maximum  value of  the streak amplitude, $U_s$
as a function of the stochastic excitation  amplitude, $\varepsilon$, for  an NL$_1$ simulation (red),  an ensemble NL$_{10}$  simulation (green),  an ensemble NL$_{100}$
simulation (blue), and an ensemble RNL$_{100}$ simulation (black).
The critical bifurcation value has been determined from stability analysis of the S3T system
and it has been confirmed that this value is closely approximated using  RNL$_{100}$.
For $\varepsilon/\varepsilon_c < 1$,  S3T predicts that  the streamwise streak and roll amplitude
is zero. At $\varepsilon=\varepsilon_c$  the S3T spanwise uniform equilibrium bifurcates giving rise to a finite amplitude  equilibrium
with  roll and streak. The NL$_1$ and NL$_{10}$  simulations  exhibit fluctuating streak/roll structures
and one standard deviation of the fluctuations correspond to the shaded regions in the figure.
The fluctuations in the ensemble NL$_{100}$ and RNL$_{100}$ simulations
are small and only those associated with NL$_{100}$ are shown.
Other parameters as in Fig. \ref{fig:grS3T}.}
\label{fig:bifur_en}
\end{figure*}


Using S3T we can find roll/streak structures that are  independent of time because their forcing derives from a converged covariance obtained from an equivalently infinite ensemble of independent realizations.  These fixed point  equilibria
are imperfectly reflected in individual realizations because fluctuations in the covariance arise due to the finiteness of the equivalent ensemble of statistically independent structures that fit in the channel.  These fluctuations in the covariance result in
 imperfect correspondence with the underlying  equilibrium  structure revealed by S3T
(cf.~\cite{Farrell-Ioannou-2003-structural,Farrell-Ioannou-2015-book}).   In order to verify that the S3T fixed point does in fact underly
the dynamics of the roll/streak structure observed in RNL and NS it
is useful to obtain solutions lying on the continuum from the single realization solution to the infinite ensemble S3T fixed point solution.
S3T dynamics is approached  by  RNL$_N$ simulations as  $N\rightarrow \infty$. The RNL$_N$
system is governed by the system of equations
\begin{subequations}
\label{eq:eRNL}
\begin{gather}
\partial_t\boldsymbol{U}+ \boldsymbol{U} \cdot \nabla \boldsymbol{U}   + \nabla P -  \Delta \boldsymbol{U}/R = - \left < \[\boldsymbol{u}' \cdot \nabla \boldsymbol{u}'\]_x \right >_N\ ,
\label{eq:eRNLm}\\
 \partial_t\boldsymbol{u}_n'+   \boldsymbol{U} \cdot \nabla \boldsymbol{u}_n' +
\boldsymbol{u}_n' \cdot \nabla \boldsymbol{U}  + \nabla p_n' -  \Delta  \boldsymbol{u}_n'/R
= \sqrt{\varepsilon} ~{\boldsymbol{f}_n}'\ ,
 \label{eq:eRNLp}\\
 \nabla \cdot \boldsymbol{U} = 0\ ,\ \ \ \nabla \cdot \boldsymbol{u}_n' = 0\ ,\ \ \ \nabla \cdot \boldsymbol{f}_n' = 0\, \label{eq:eNSdiv0}
\end{gather}\label{eq:NSE0}\end{subequations}
where $n=1,\cdots,N$ indicates the ensemble member, and $\left < \cdot \right >_N $ indicates an average over the $N$ ensemble members.
Note that in correspondence with S3T dynamics the perturbation-perturbation interaction
in \eqref{eq:NSp} is ignored.

In a similar manner we can define ensemble NL$_N$ simulations
correspond  to the first two components of a converged expansion in cumulants.  NL$_N$ is governed by the system of equations:
\begin{subequations}
\label{eq:eDNS}
\begin{gather}
\partial_t\boldsymbol{U}+ \boldsymbol{U} \cdot \nabla \boldsymbol{U}   + \nabla P -  \Delta \boldsymbol{U}/R = - \left < \[\boldsymbol{u}' \cdot \nabla \boldsymbol{u}'\]_x \right >_N\ ,
\label{eq:eDNSm}\\
 \partial_t\boldsymbol{u}_n'+   \boldsymbol{U} \cdot \nabla \boldsymbol{u}_n' +
\boldsymbol{u}_n' \cdot \nabla \boldsymbol{U}  + \nabla p_n' -  \Delta  \boldsymbol{u}_n'/R
=\nonumber\\\hspace{4em}=  - \(  \boldsymbol{u}_n' \cdot \nabla \boldsymbol{u}_n' - \[\boldsymbol{u}_n' \cdot \nabla \boldsymbol{u}_n'\]_x \,\)+
\sqrt{\varepsilon} ~{\boldsymbol{f}_n}'\ ,
 \label{eq:eDNSp}\\
 \nabla \cdot \boldsymbol{U} = 0\ ,\ \ \ \nabla \cdot \boldsymbol{u}_n' = 0\ ,\ \ \ \nabla \cdot \boldsymbol{f}_n' = 0\, \label{eq:NSdiv0}
\end{gather}\label{eq:eNSE00}\end{subequations}

We are interested in whether the analytical predictions of the S3T equations are approached in RNL$_N$ and NL$_N$ simulations as $N$ increases.
Results are presented for  the minimal Couette flow channel of  Hamilton, Kim \& Waleffe~ \cite{Hamilton-etal-1995} at $R=400$ (based on channel half-width)
with streamwise length
$L_x = 1.75 \upi$, spanwise length $L_z = 1.2 \upi$  and  channel half-width $L_y=1$.
The gravest streamwise wavenumber $k_x = 2 \upi / L_x$ is stochastically excited  using
independent compact support cross-stream velocity and cross-stream
vorticity structures in  $(y, z)$. Numerical calculations  employ $N_y=21$ grid points in the cross-stream direction and
$32$ harmonics in the spanwise and streamwise directions. Other stochastic excitations
produce only qualitative differences in the results.
A study of the  S3T dynamics of  this channel model was reported in \cite{Farrell-Ioannou-2012}.

  \begin{figure}
 \includegraphics[width = \columnwidth]{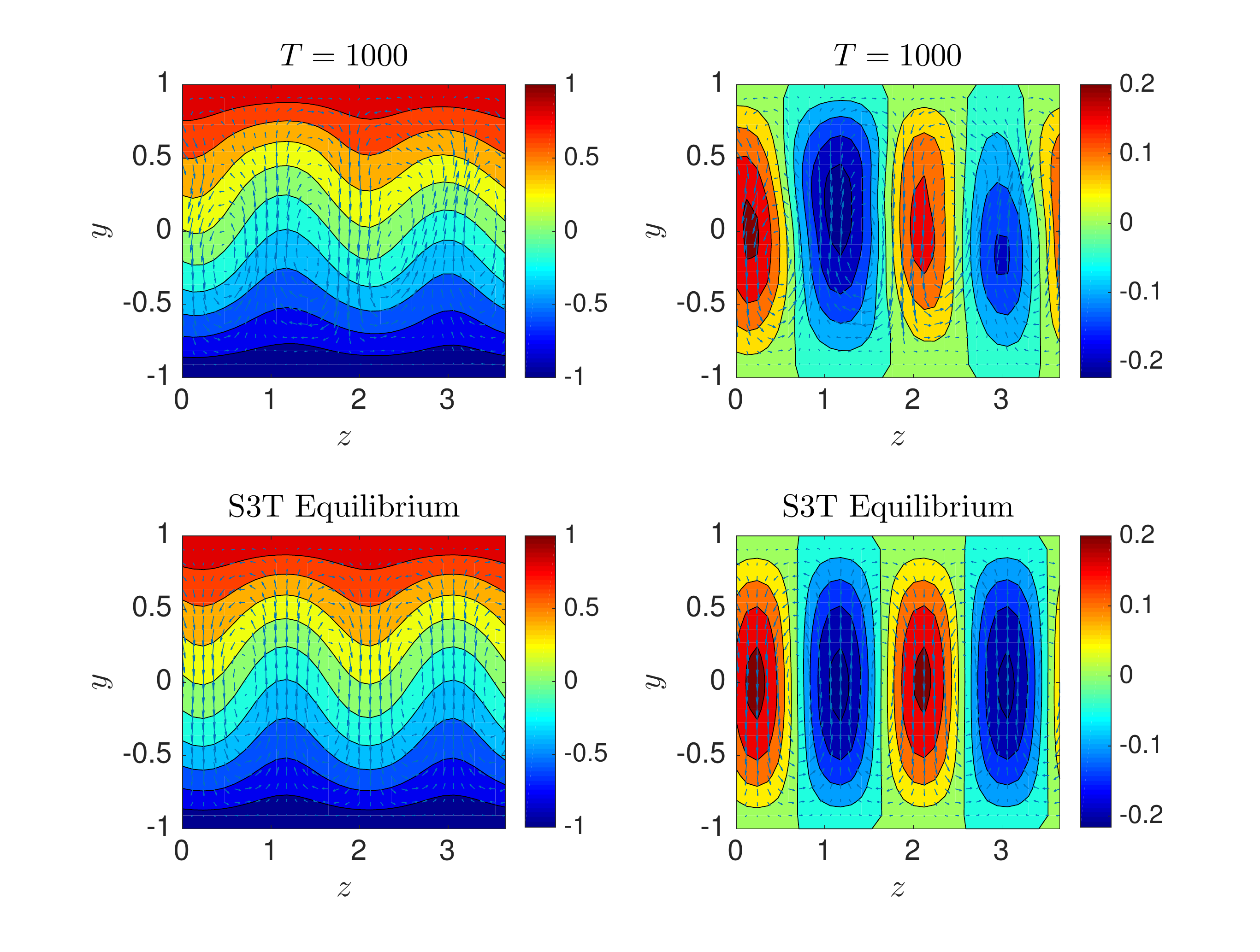}
 \caption{Top panels: Snapshot of the streamwise mean flow from an NL$_1$ simulation at stochastic  excitation amplitude $\varepsilon/\varepsilon_c=3$. Shown are contours of the  streamwise mean velocity $U$  (left top), streak velocity, $U_s$ (right top)  and velocity vectors of the components $(V,W)$  in the
$(y,z)$ plane  at $t=1000$ of the simulation.
Bottom panels: The corresponding streamwise mean flow for the S3T system at    $\varepsilon/\varepsilon_c=3$.
This figure shows that the   equilibrium roll/streak regime
predicted by S3T is reflected in single realizations of the NL equations. The development
of the roll/streak structure in a NL$_1$ simulation can be seen in MOVIE1 (cf. supplementary materials).
The development
of the roll/streak equilibrium   in a S3T equilibrium simulation can be seen in MOVIE2 (cf. supplementary materials).
 Parameters are as in the previous figures.}
\label{fig:state_f20}
\end{figure}

 \begin{figure}
 \includegraphics[width = \columnwidth]{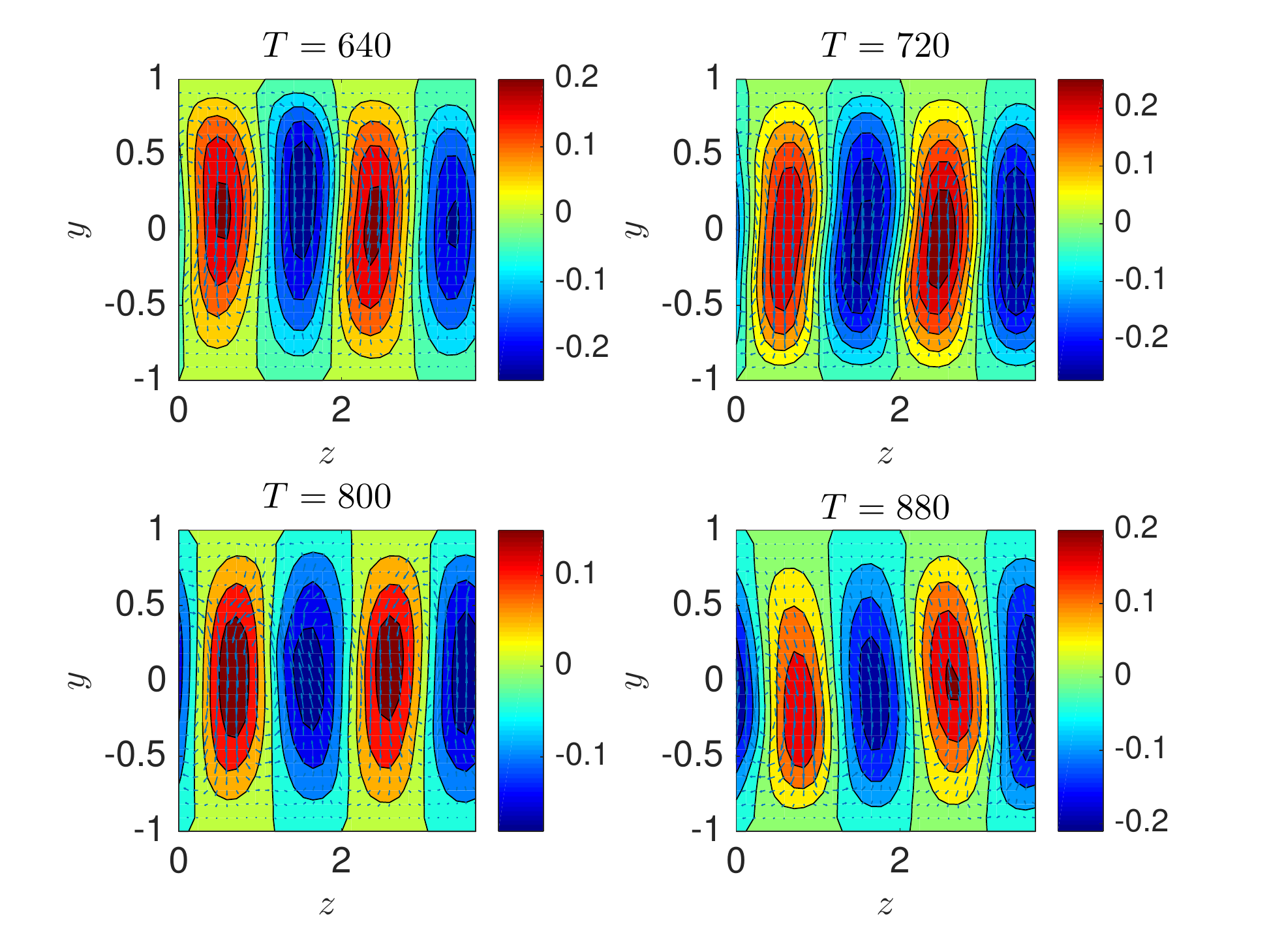}
\caption{Snapshots at times $t=640, ~720,~ 800,~880$ of the contours of the streak velocity, $U_s$, and velocity vectors of the components $(V,W)$  plotted on a $(y,z)$ plane cross-section
from an NL$_1$ simulation at stochastic excitation amplitude  $\varepsilon/\varepsilon_c=3$.  This figure shows the persistence of the organized structure
in NL$_1$. This structure and its persistence stem from the underlying equilibrium state that
exists for this excitation amplitude in the S3T dynamics. The other parameters are as in the previous figures.}
\label{fig:4panels}
\end{figure}

\section{Comparison of roll/streak bifurcation and structure in S3T, RNL$_N$ and DNS$_N$}
\label{sec:framework}

The S3T SSD  \eqref{eq:S3T}   supports spanwise uniform fixed point  solutions with streamwise mean flow form  $\U_e= (U_{e}(y),0,0)$ and associated  spanwise  covariance $C_{e}(y_1,y_2,z_1-z_2)$.
Taking  $\varepsilon=0$,   recovers the  laminar  Couette flow $U_e=y$  with $C_e=0$.  As $\varepsilon$ increases the equilibrium streamwise mean flow, $U_e(y)$, departs
from the Couette flow.    Stability of these spanwise uniform S3T equilibria  can be determined as a function of $\varepsilon$ using the S3T  equations \eqref{eq:S3T} linearized about these fixed points  \citep{Farrell-Ioannou-2012}.

Eigenvalues and the associated mean flow eigenfunction structure for the first two most unstable S3T modes are shown in
 Fig. \ref{fig:grS3T}. The  complete associated eigenfunctions  comprise both a mean flow component $(\delta U(y,z), \delta V(y,z), \delta W(y,z))$,
 which is shown in Fig. \ref{fig:grS3T},
 and  a covariance component
 $\delta C(y_1,y_2,z_1,z_2)$.   The structure of the mean flow component of these eigenfunctions changes only slightly  as the amplitude of the forcing, $\varepsilon$, increases.   The eigenfunctions
 consist of low and high speed streamwise streaks together with roll circulations exactly collocated  to reinforce the streak velocity.
Despite being more highly dissipated by diffusion,  the mean flow eigenfunction which
becomes unstable first as $\varepsilon$ increases is not the eigenstructure with the gravest  spanwise wanumber $k_z= 2\upi/L_z =1.67$, \ shown in Fig. \ref{fig:grS3T}b, but the second spanwise harmonic with wavenumber $k_z= 4 \upi/L_z = 3.33$, shown in Fig. \ref{fig:grS3T}a.   Destabilization
of these roll/streak eigenfunctions can be traced to a universal positive feedback mechanism operating in turbulent flows:  when  incoherent turbulence is perturbed by a coherent streak, the streak distorts the incoherent turbulence so as to induce ensemble mean  Reynolds stresses forcing streamwise mean roll circulations configured to reinforce the streak perturbation that gave rise to them (cf. \cite{Farrell-Ioannou-2012}).  The modal streak perturbations of the fastest growing eigenfunctions  induce the strongest such feedback (when account has been taken for viscous damping).

{We note that neutral mode and critical layer based self-sustaining process (SSP) theories
such as the vortex-wave interaction theory (VWI)  predict structures at variance with the S3T
unstable modes we obtain. As shown in  Fig. \ref{fig:grS3T}a,b  organization of
the Reynolds stress by the streak (even at perturbational amplitude) results in a smooth domain-wide
forcing of the roll circulation.  At high Reynolds number in the neutral mode SSP and VWI mechanisms
this interaction  is localized at the critical layer
\citep{Hall-Smith-1991,Deguchi-Hall-2014,Hall-Sherwin-2010,Wang-etal-2007}.
The interaction between  perturbations and mean flow   in neutral mode and VWI theories by
necessity occurs near the the critical layer  in the inviscid limit
because according to the non-acceeleration theorem  at steady state and in the absence of forcing and dissipation there is no interaction between mean and perturbations except at the critical layer  \citep{Eliassen-Palm-1961,Boyd-1976,Andrews-McIntyre-1976-I}. In S3T there is forcing and consequently the interaction  is not required to be concentrated in the vicinity of the critical layer.
The modes we calculate organize distributions of
Reynolds stress with divergence exactly coherent with the mode roll structure, as is required of
a mode solution, and not in any sense concentrated at a critical surface. In fact the lack of any
evidence for concentration of Reynolds stress divergence at a particular cross-stream location
either in our stable roll/streak regime or in our self-sustaining turbulence
simulations argues against a mechanism relying on an interaction localized
at a critical surface.}

S3T stability analysis determines the critical excitation, $\varepsilon_c$, at which  the spanwise homogeneous turbulent equilibrium state becomes unstable.  For the parameters of our example problem this  $\varepsilon_c$ corresponds to maintaining in the Couette flow a perturbation field with mean energy density  $0.14 \%$ of the energy density of the
Couette flow.
For $\varepsilon > \varepsilon_c$   a symmetry breaking occurs with the emergence of mean flow structures in the form of the fastest growing eigenfunction  which is shown in
Fig. \ref{fig:grS3T}a.  Over a finite interval  $\varepsilon_c < \varepsilon < \varepsilon_t$ the unstable S3T eigenfunction  equilibrates nonlinearly to
form  finite amplitude S3T equilibria with  roll/streak structure qualitatively similar to the
corresponding eigenfunction  (for our  examples $\varepsilon_t/\varepsilon_c \approx 5.5$).

A bifurcation diagram showing  the maximum of the streak velocity, $U_s$, and of the streamwise mean cross-stream velocity, $V$,
is shown as  a function of $\epsilon$  in Fig. \ref{fig:bifur_en} .  The indicated critical $\epsilon_c$ was determined by S3T stability analysis.
For $\varepsilon/\varepsilon_c<1 $ the equilibrium is spanwise independent with no coherent roll/streak structure.  The  equilibrium  values shown
in  Fig. \ref{fig:bifur_en}  were obtained using  RNL$_{100}$ simulations.  These   RNL$_{100}$  equilibria have been verified to be very close  to the infinite ensemble  S3T equilibria.

Single  NL and ensemble NL integrations allow us to study the correspondence between  the infinite ensemble predictions of S3T analysis and NL turbulence.  While finite ensemble simulations produce fluctuating roll/streak structures we find that even in the case of a realization simulation, corresponding to $N=1$,   a clear roll/streak structure emerges  for
$\varepsilon > \varepsilon_c$  which exhibits  great persistence  and has  the same structure as  that predicted by S3T analysis.
An indicative comparison between an S3T equilibrium roll/streak structure and a snapshot of the corresponding roll/streak from an NL$_1$ simulation at $\varepsilon/\varepsilon_c=3$ is shown  in Fig. \ref{fig:state_f20}.

While the S3T equilibria are fixed points, the corresponding roll/streak
structure in  the NL$_1$  simulation reflect the time independence of the S3T equilibria imperfectly.  However,  it is persuasive that the analytical structure revealed by S3T analysis underlies the behavior seen in the NL$_1$ simulation; for example see the snapshots shown in
Fig. \ref{fig:4panels}.
Noise driven fluctuations of the ensemble structure  are also apparent in the bifurcation diagram shown in Fig. \ref{fig:bifur_en} in which the
mean and   variance of  the maximum streak, $U_s$,    in NL$_{1}$ and NL$_{10}$ are indicated.   The reflection of the analytical S3T  bifurcation is clearly seen in the NL$_{10}$ results and  near convergence is obtained  in the NL$_{100}$ results.

 \begin{figure}
 \includegraphics[width = \columnwidth]{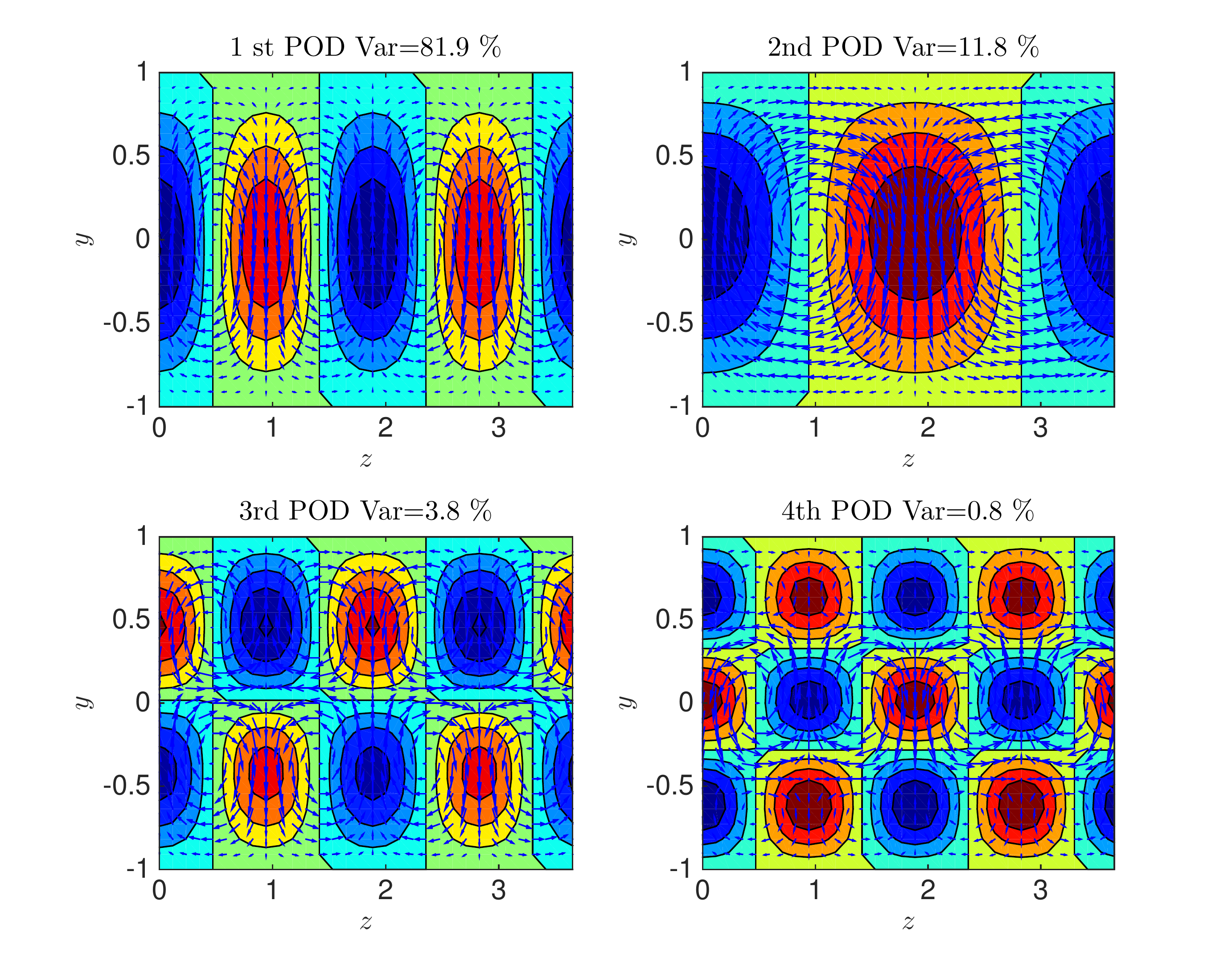}
 \caption{Contours of  streak velocity, $U_s$, and vectors of roll components $(V,W)$  plotted on a $(y,z)$  cross-section
for  the  first 4 PODs of the streamwise mean flow  fluctuations of an NL$_1$ forced at $\varepsilon/\varepsilon_c= 0.75 $.
The PODs come in pairs. The first pair of PODs which account for $82 \%$ of the energy of the fluctuations of the streamwise mean flow
has the structure of the least damped S3T mode which because of the  synergistic mechanism revealed by S3T  is not the gravest mode in the channel.
This figure shows that the fluctuations in the NL$_1$ simulations
reveal the S3T stable modes.  Other parameters as in the previous figures.}
\label{fig:PODs_f5}
\end{figure}

 \begin{figure}
 \includegraphics[width = \columnwidth]{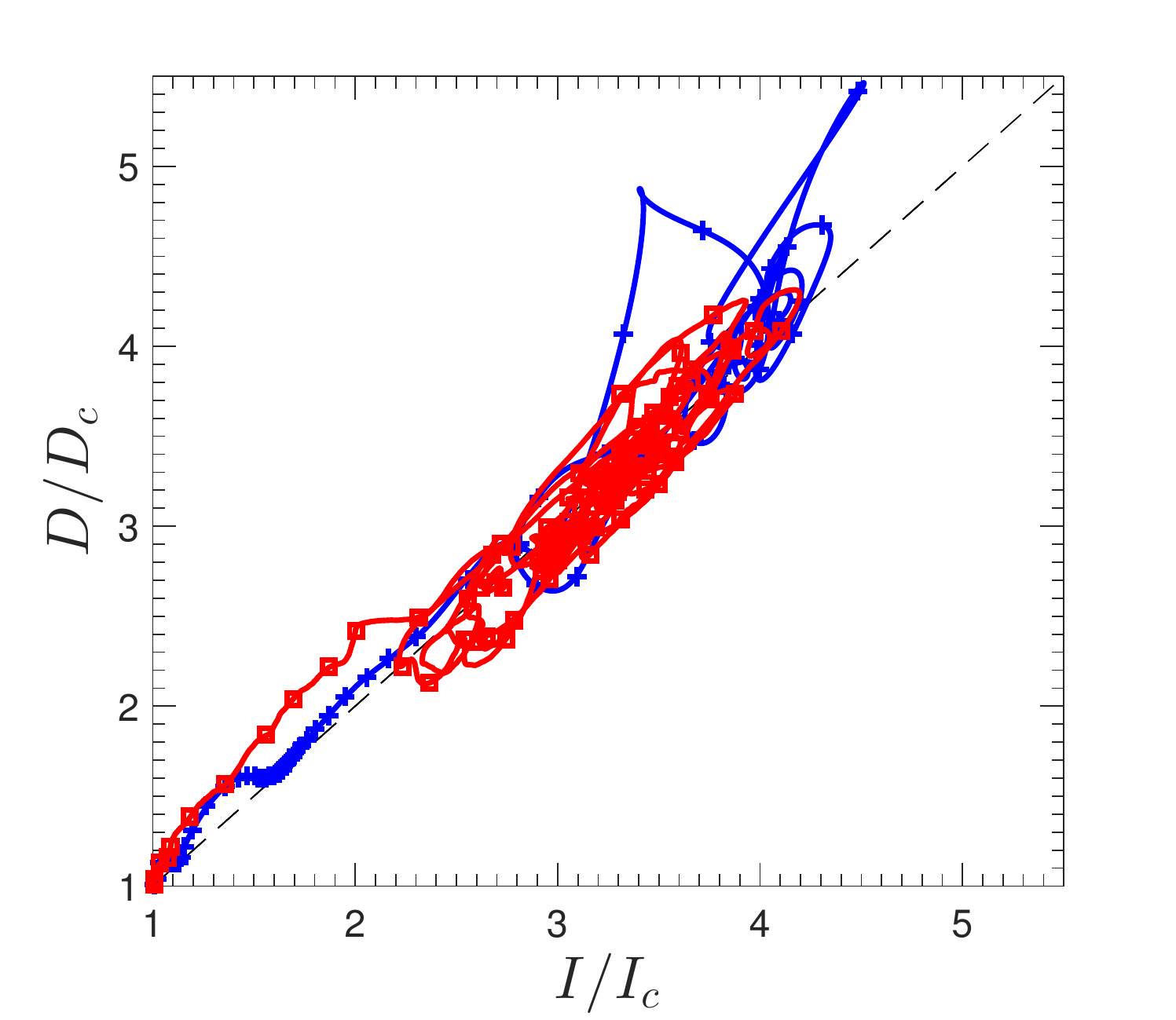}
 \caption{Evolution of energy input rate, $I/I_c$ and dissipation rate, $D/D_c$, from the laminar state to the turbulent state in   an NL$_1$ simulation   (squares-solid) and
in  an S3T simulation  (crosses-solid) with  background turbulence excitation parameter $\varepsilon / \varepsilon_c=9$. 
 Symbols are marking intervals of 10 units of time. The metastable state is characterized by $D/D_c \approx 1.7$.
 Parameters as in the previous figures.}
\label{fig:ID_f60}
\end{figure}

We have demonstrated that the unstable roll/streak modes and associated finite amplitude S3T
equilibria  that are revealed by S3T analysis give rise to  the structure observed in pre-transitional turbulent
Couette flow in both NL and ensemble NL simulations.  However, the
stable S3T modes supported in the
S3T stable the interval ($0<\varepsilon/\varepsilon_c <1$) are also important structures in the dynamics of pre-transitional turbulence.   While not excited in the fluctuation free S3T dynamics, these stable S3T modes are robustly  excited by  fluctuations
in the forcing in NL$_1$ simulations  (cf. \citep{Farrell-Ioannou-2003-structural,Farrell-Ioannou-2015-book,Constantinou-etal-2014}).
Correspondingly, for subcritical excitation ($0<\varepsilon/\varepsilon_c <1$) the mean flow of NL  or ensemble NL simulations reveals a ubiquitous tendency to form roll/streak structures with temporally variable $(y,z)$ structure  arising from excitation of the stable manifold of S3T eigenmodes.
A POD analysis (cf. \cite{Berkooz-etal-1993}) of the streamwise mean flow reveals the dominance of this component
of the variability which is accounted for by excitation of these roll/streak structures predicted by S3T (cf. \cite{Nikolaidis-etal-Madrid-2016}).   For example, the first 4 POD's of
NL$_1$ at $\varepsilon/\varepsilon_c=0.75$, shown in Fig. \ref{fig:PODs_f5},
have the structure predicted by the S3T eigenmodes.  Consistent with S3T analysis the first POD  corresponds to the mode with spanwise wavenumber
$k_z=4 \upi/L_z$, which corresponds to the least stable eigenfunction at this $\varepsilon/\varepsilon_c$.  Note
that  all POD's exhibit exact alignment of the roll circulations with the streaks. This provides confirmation of the S3T prediction that these are the modal structures predicted by S3T.
Consistent with these stable modes being  excited  by turbulent fluctuations,  as $\varepsilon / \varepsilon_c \rightarrow 1$   fluctuations of roll/streak form exhibit enhanced variance
(cf. Fig. \ref{fig:bifur_en})  which is indicative of approach to a bifurcation and is a phenomenon analogous to  that of critical opalescence on approach to a fluid phase transition.

 \begin{figure}
 \includegraphics[width = \columnwidth]{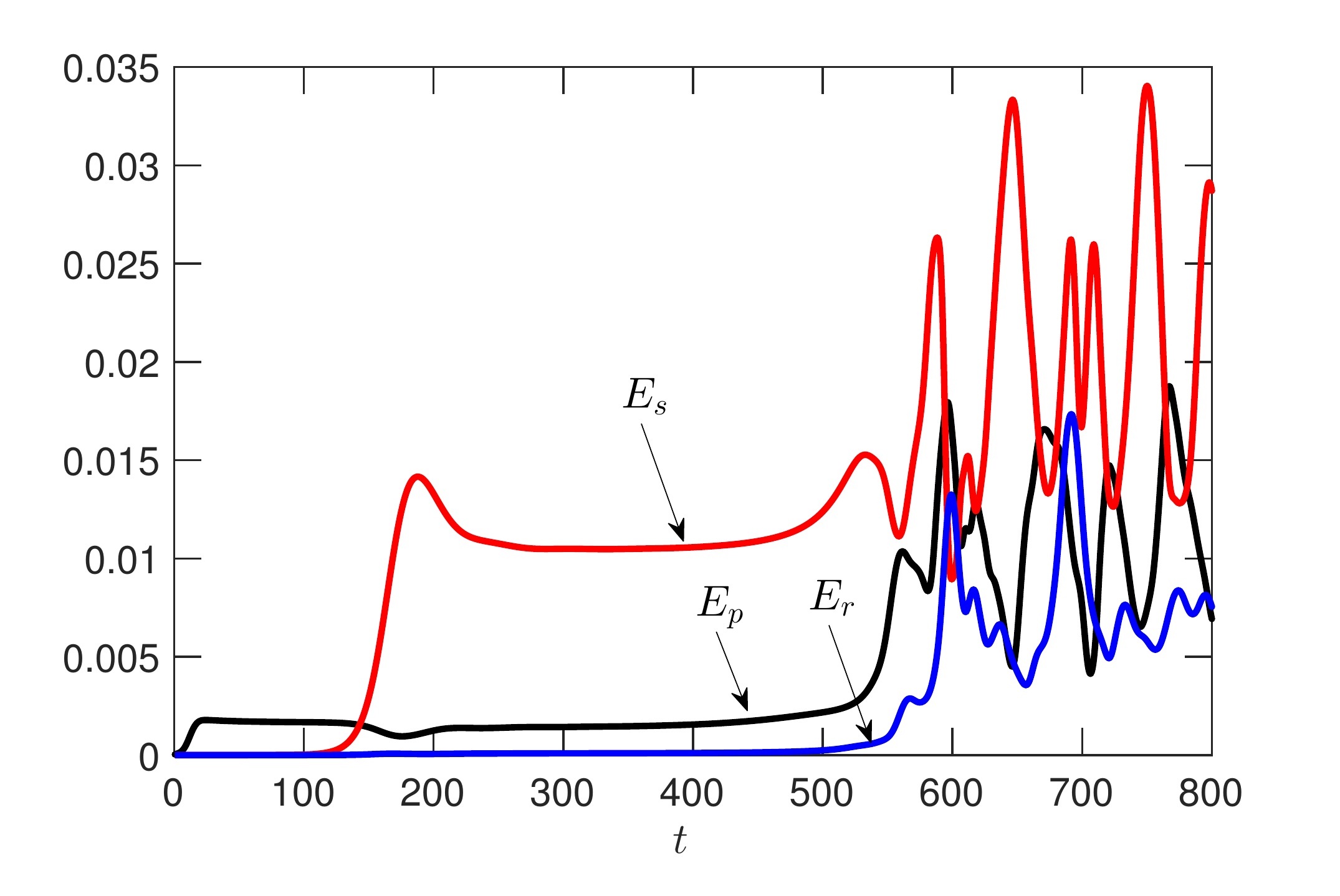}
\caption{Evolution of the streak energy, $E_r$, roll energy, $E_r$, and perturbation energy
$E_p$, in an S3T integration at $\varepsilon/\varepsilon_c= 9$ under spanwise homogeneous forcing.
The flow is initialized   with
a small random streamwise  mean perturbation  with spanwise dependence  in order to break  spanwise symmetry. The spanwise symmetric S3T equilibrium is unstable
and a  quasi-steady state emerges by time $t=200$  with the roll/streak  structure shown in Fig. \ref{fig:states_f60}. At this supercriticality
the roll/streak structure (cf. Fig. \ref{fig:unstable_f60}) is an unstable fixed point  of the S3T dynamics
and the flow transitions to the turbulent state.
Other parameters as in the previous figures.}
\label{fig:e_f60}
\end{figure}

\begin{figure*}
	\centering
       \includegraphics[width = .8\textwidth]{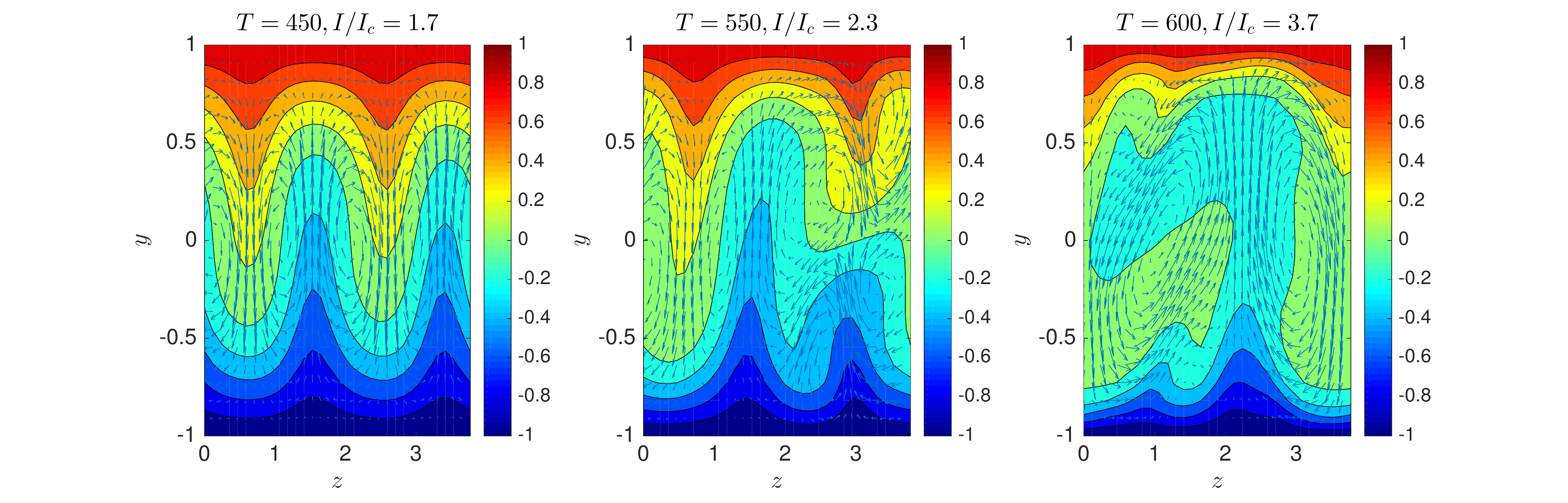}
       \caption{ Snapshots of the streamwise mean flow as it undergoes S3T  transition to turbulence under stochastic forcing.
Shown are contours of the streak velocity, $U_s$, and velocity vectors of the components $(V,W)$  plotted on the $(y,z)$ plane.
A quasi-steady roll/streak is first formed  (cf. left panel) with
with input energy rate $I/I_c \approx 1.7$
and the structure of the fastest growing
S3T instability (cf. Fig. \ref{fig:grS3T}) which has spanwise wavenumber $k_z=4 \upi/L_z$.  At about $t=550$ the flow transitions  through oscillations
to  a turbulent roll/streak  with a  dominant  $k_z=2 \upi/L_z$ structure. The transition period can be
extended by enforcing the mirror symmetry of the streak-roll structure about the streak maximum.
 Other parameters as in the previous figures.}
\label{fig:states_f60}
\end{figure*}


\section{Transition to turbulence}

At background turbulence excitation parameters exceeding $\varepsilon_t$ ($\varepsilon_t / \varepsilon_c \approx 5.5$  for the chosen parameters)
the finite amplitude roll/streak
equilibria are no longer S3T stable and the flow transitions to a turbulent state,
which is self-sustaining and persists even when the background turbulence excitation parameter is subsequently set
to $\varepsilon=0$ (cf. \cite{Farrell-Ioannou-2012}).
RNL$_1$ and NL$_1$ also transition to essentially similar self-sustaining turbulence.
Example trajectories of transition from the laminar  equilibrium state to the turbulent attractor  for  NL$_1$ and S3T are
 shown in Fig. \ref{fig:ID_f60}.

\begin{figure}
 \includegraphics[width = \columnwidth]{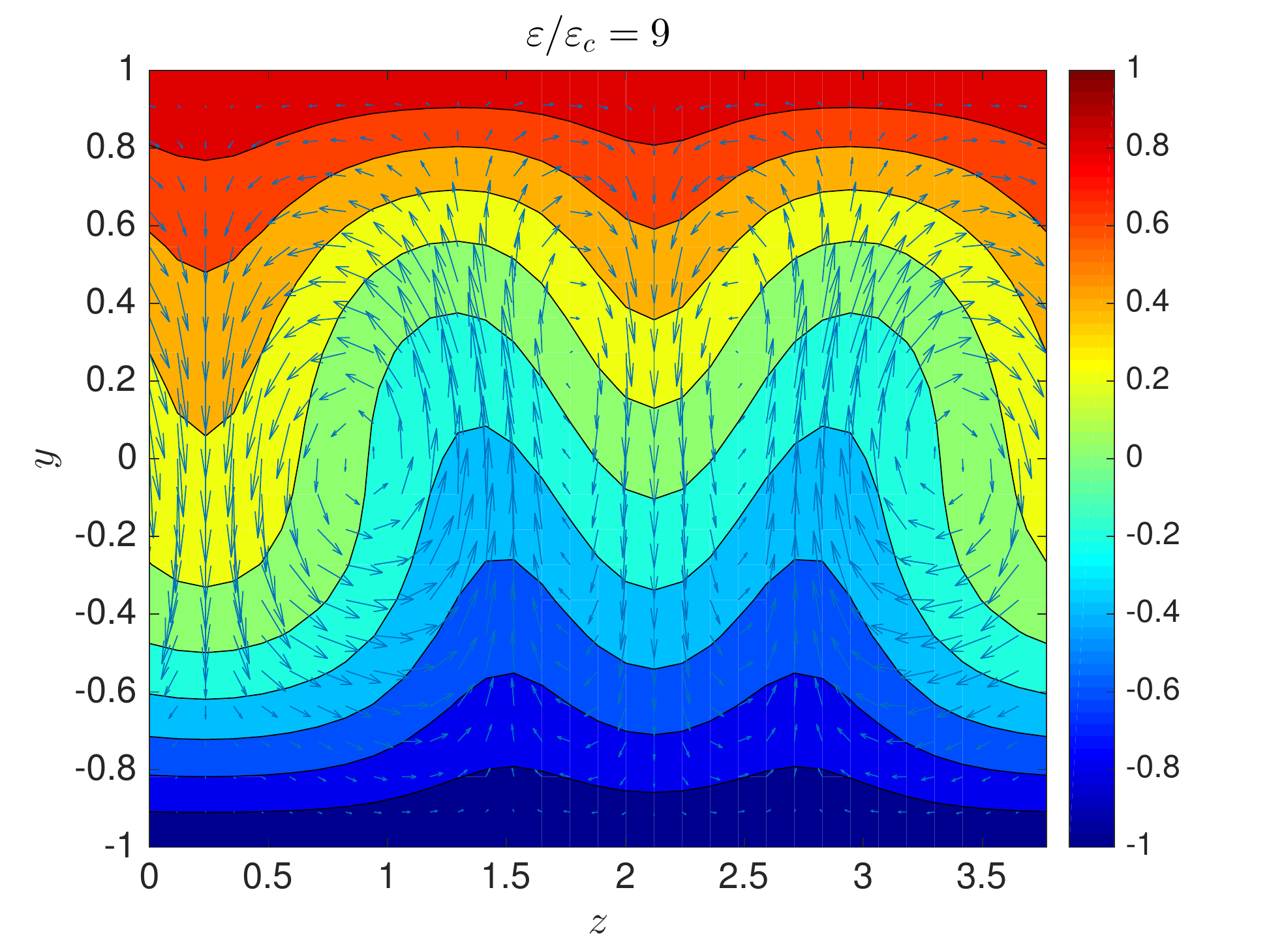}
\caption{ The unstable roll/streak S3T equilibrium at $\varepsilon/\varepsilon_c= 9 $.   Shown are
contours of the streak velocity, $U_s$, and velocity vectors of the components $(V,W)$  plotted on a $(y,z)$ plane cross-section.
 Other parameters as in the previous figures.}
\label{fig:unstable_f60}
\end{figure}

A typical evolution  of the perturbation energy density, $E_p$,
streak energy density, $E_s$, and roll energy density, $E_r$, of background turbulence excitation parameter $\varepsilon / \varepsilon_c = 9$
is shown in Fig. \ref{fig:e_f60}  for the case of S3T.
The S3T integration was initialized with  a small random
streak perturbation.  The flow transitions to turbulence at time $T \approx 550$.
In this transition  process the roll/streak   emerges at first as an S3T instability
which equilibrates by time
$T\approx 200$ to the quasi-equilibrium finite amplitude
roll/streak structure   shown in the left panel of Fig. \ref{fig:states_f60}.
This quasi-equilibrium is associated with an energy
input-rate $I/I_c \approx 1.7$,  which lies approximately midway between the value associated with the laminar state and that associated
with the statistical mean of the turbulent state.
At these parameters there exists near this quasi-equilibrium a symmetric unstable equilibrium,
shown in  Fig. \ref{fig:unstable_f60}, which can be converged to by suppressing spanwise asymmetries.
The  roll/streak structure that emerged in the S3T in the presence of realistic spanwise asymmetric perturbations
breaks  by exciting the unstable directions of the unstable equilibrium   at about $T \approx 550$ and the flow transitions to turbulence.
While this pathway to turbulence is typical in all S3T simulations with $\varepsilon > \varepsilon_t$ the timing of transition depends on
the structure of the initialized state  which determines the projection on the instability of the S3T equilibrium state.
 For example, if the flow state at $\varepsilon / \varepsilon_c=9$
is constrained to have no perturbations breaking mirror-symmetry in the spanwise direction
the flow equilibrates to the unstable roll/streak structure shown in  Fig. \ref{fig:unstable_f60} without
ever transitioning to turbulence, while if the initial flow state includes a rich spectrum of such perturbations
the  meta-stable period is appreciably shortened.

\begin{figure*}
	\centering
       \includegraphics[width = .8\textwidth]{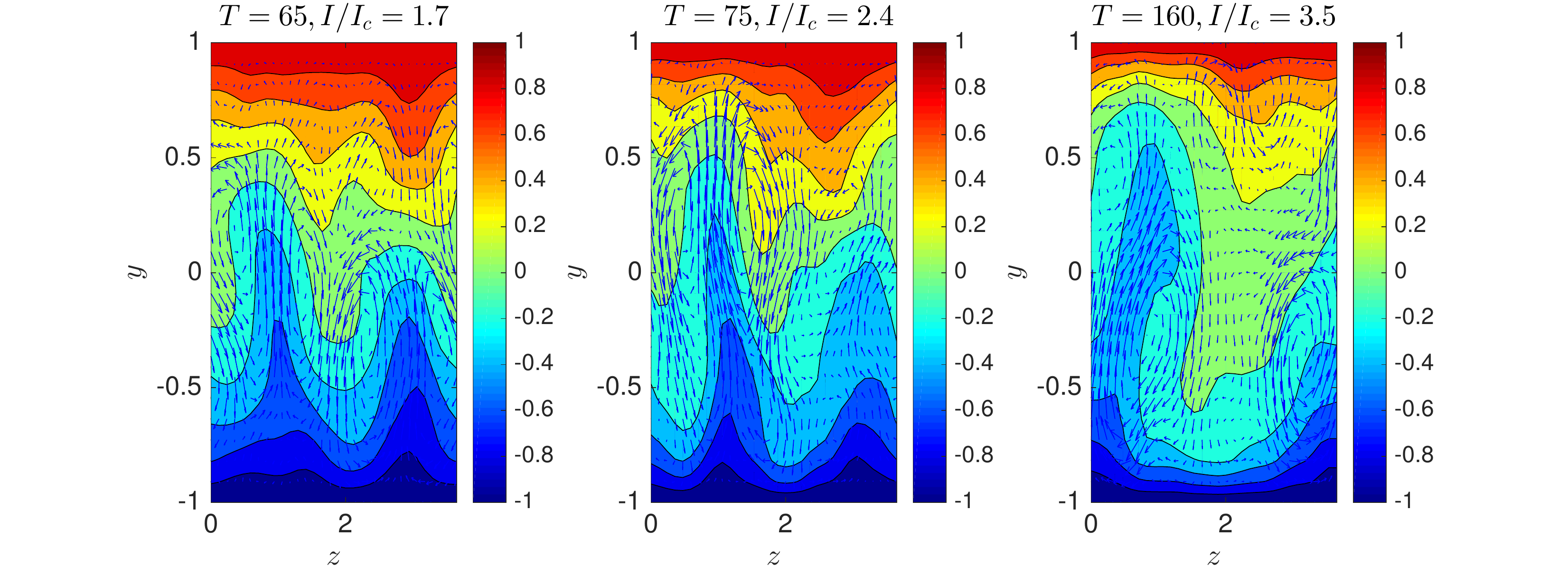}
       \caption{ Snapshots of the streamwise mean flow as it undergoes   transition to turbulence in a NL$_1$ simulation under stochastic forcing.
with $\varepsilon/\varepsilon_c= 9$. Shown are contours of the streak velocity, $U_s$, and velocity vectors of the components $(V,W)$  in the $(y,z)$ plane.
A quasi-steady roll/streak initially forms, by $T=65$, that swiftly  breaks  down and the flow transitions to turbulence.
The transition is as in the S3T simulation (cf.  Fig. \ref{fig:e_f60} and Fig. \ref{fig:states_f60}),
except that the flow passes through the metastable state  rapidly.
 Other parameters as in the previous figures.}
\label{fig:states_dns_f60}
\end{figure*}

This sequence of events, with  rapid break-down of the finite amplitude roll/streak structure,
is observed in NL$_1$ simulations at $\varepsilon / \varepsilon_c=9$ when the simulation is  initialized with the laminar state.
The roll/streak structure associated with the underlying S3T instability  arises at first, as in the S3T
simulation, but then rapidly
transitions to the turbulent state.
 Snapshots of the roll/streak structure during this transition, which occurs by $T  =90$,  are
 shown in Fig. \ref{fig:states_dns_f60}.

%

\section{Conclusion}

SSD makes available to analysis the manifold of nonlinear instabilities associated with the systematic organization of the
background turbulence by coherent structures.  In this work the S3T implementation of  SSD was used to  study instabilities of this type and their nonlinear extensions in a minimal channel configuration of Couette flow.
At first a manifold of stable modes with roll/streak form is supported as the parameter controlling
the background turbulence intensity, $\varepsilon$,  is increased from zero.
The least stable mode of this manifold is destabilized at a critical excitation
designated $\varepsilon_c$ and a  finite amplitude stable fixed point with roll/streak structure arises for excitations between
$\varepsilon_c$ and  a second critical value for which the finite amplitude equilibrium roll/streak is destabilized, designated
$\varepsilon_t$.   For excitation exceeding   $\varepsilon_t$
the roll/streak equilibrium is unstable to spanwise asymmetric perturbations and
becomes time-dependent  resulting in the establishment of the turbulent state with spanwise wavenumber approximately half that
of the equilibrium state.  This sequence of states and transitions suggests a route to  turbulence  in a developing boundary layer.
In order to study these SSD states and their dynamics in more detail their correspondence to realization dynamics was examined
making use of a comparison  among the predictions of S3T and ensemble implementations of a quasi-linear model sharing the
dynamical restrictions of S3T (RNL$_N$) and the associated nonlinear model (NL$_N$).  Although the SSD instabilities
and their associated fixed point nonlinear equilibria and time dependent statistical mean attractor states have analytical expression only in the S3T implementation of the equivalently infinite ensemble SSD dynamics, the predicted  dynamics is clearly reflected  in both the dynamically similar
quasi-linear system (RNL$_1$) and in DNS (NL$_1$).    This correspondence was further examined  using ensemble
implementations of the RNL and DNS
systems.   As a consequence of sharing the same dynamical restrictions, the
RNL$_N$ system converges to S3T an $N \rightarrow \infty$.  Remarkably, the NL$_N$ system, which corresponds to a full
closure for this problem, also converges to close correspondence with S3T as $N \rightarrow \infty$.  This convergence is reflected
in similar bifurcation behavior  as well as similar equilibrium structures for the stable fixed point equilibria.
Additionally, S3T also predicts a second bifurcation at a higher value of the turbulent excitation
parameter that results in destabilization of the finite amplitude roll/streak equilibria and establishment of a turbulent
state corresponding to minimal channel turbulence.  This scenario constitutes a mechanism for bypass
transition to the turbulent state.  Comparison with NL$_1$ reveals that this mechanism in fact  is
responsible for bypass transition in the case that the  transition is instigated by background turbulence rather than by an
optimal perturbation imposed at sufficiently high amplitude.

\begin{acknowledgments}
Brian Farrell was partially supported by  NSF AGS-1246929. We thank Daniel Chung, Navid Constantinou, Dennice Gayme and
Vaughan Thomas for helpful discussions.
\end{acknowledgments}

%

\end{document}